\documentclass[prb,preprint,superscriptaddress]{revtex4}
\usepackage{amsmath}
\usepackage{amssymb}
\usepackage{amsbsy}
\usepackage{graphicx}
\newcommand{\unit}[2]%
{\mbox{\ensuremath{#1}}\mbox{\,\ensuremath{\mathrm{#2}}}}
\newcommand{\Tc}{\ensuremath{T_\mathrm{c}}}
\newcommand{\Jc}{\ensuremath{J_\mathrm{c}}}
\newcommand{\Ic}{\ensuremath{I_\mathrm{c}}}
\newcommand{\JcG}{\ensuremath{J_\mathrm{c,G}}}
\newcommand{\JcGB}{\ensuremath{J_\mathrm{c,GB}}}
\newcommand{\JcB}{\ensuremath{\Jc(B)}}

\newcommand{\mrem}{\ensuremath{m_\mathrm{rem}}}
\newcommand{\Hmax}{\ensuremath{H_\mathrm{max}}}

\newcommand{\Birr}{\ensuremath{B_\mathrm{irr}}}

\begin{document}
\title{Neutron irradiation of coated conductors}
\author{Eisterer M}
\email{eisterer@ati.ac.at}
\author{Fuger R}
\author{Chudy M}
\author{Hengstberger F}
\author{Weber H W}
\affiliation{Atominstitut, Vienna University of Technology,
Stadionallee 2, 1020 Vienna}


\begin{abstract}
Various commercial coated conductors were irradiated with fast
neutrons in order to introduce randomly distributed, uncorrelated
defects which increase the critical current density, \Jc, in a
wide temperature and field range. The \Jc-anisotropy is
significantly reduced and the angular dependence of \Jc\ does not
obey the anisotropic scaling approach. These defects enhance the
irreversibility line in not fully optimized tapes, but they do not
in state-of-the-art conductors. Neutron irradiation provides a
clear distinction between the low field region, where \Jc\ is
limited by the grain boundaries, and the high field region, where
depinning leads to dissipation.

\end{abstract}
\maketitle
\section{Introduction}
Neutron irradiation is certainly not practical for improving the
properties of long length commercial superconductors, but a very
efficient tool for benchmarking and investigating flux pinning. It
helps understanding vortex physics \cite{Mei85,Wer00,Zeh04} and
limitations of the current flow \cite{Ton02,Eis06,Eis07}.
Efficient pinning centers are introduced, whose size perfectly
matches the coherence length of YBa$_2$Cu$_3$O$_{7-\delta}$
(YBCO). They improve the critical currents, if the density of
defects was originally too small, but diminish them, if the total
defect concentration becomes too high. Although the resulting
defect structure is not perfectly well defined, since only the
density of the largest defects, the collision cascades, is known
\cite{Fri93} and many smaller defects of unknown density are
produced \cite{Sau98}, it is clearly established that all defects
are randomly distributed and uncorrelated, i.e., without a
preferred orientation. It is particularly interesting to study the
interplay of these defects with the correlated disorder usually
present in coated conductors or thin films
\cite{Fol07,Gut07,Che09}. In the present study, the effect of
neutron irratiation on commercial coated conductors was
investigated. Similarities and differences in differently produced
tapes will be presented in the following in order to provide a
comprehensive picture of neutron irradiation of this class of
materials. We do not aim at comparing the properties of the
unirradiated samples, since they were supplied at different times
and the rapid development during the past few years would not
allow a useful comparison between the different production routes.
The different stages of optimisation turned out to be the main
source of differences in the radiation response of the
investigated tapes.
\section{Experimental}
Three different types of commercial YBCO tapes were studied. The
first type of samples (denoted as MODRAB in the following) was
prepared by metal organic deposition (MOD) on a RABITS template.
The second series of samples (PLDYSZ) is based on an IBAD (ion
beam assisted deposition) YSZ (yttrium stabilized zirconia)
template. The superconducting layer was prepared by pulsed laser
deposition (PLD). The third series (MOCVDMgO) had IBAD MgO
templates with the YBCO layer made by metal organic chemical vapor
deposition (MOCVD). The MODRAB samples were irradiated to higher
fluences ($4\times 10^{21}$\,m$^{-2}$ and $10^{22}$\,m$^{-2}$)
than the PLDYSZ and MOCVDMgO samples ($2\times
10^{21}$\,m$^{-2}$). All samples were characterized by four probe
transport prior to and after irradiation by fast neutrons. The
transport measurements were made under helium gas flow up to
fields of 15\,T. The samples were measured in different fields and
temperatures and also rotated in the magnetic field under the
maximum Lorentz force configuration ($J\perp H$). The current was
continuously ramped until an abort criterion was reached. The
critical current was defined by the 1\,$\mu$Vcm$^{-1}$ criterion.
Indium pressed contacts are most suitable for our experiments
because of their low resistivity, reliability and removability,
which is important for the irradiation process.  The transition
temperature (\Tc) and the irreversibility lines were measured with
a constant current of 10\,mA, while the temperature was slowly
decreased at fixed applied fields. \Tc\ and $T_\mathrm{irr}$ were
evaluated at 0.1\,$\mu$Vcm$^{-1}$. Only the data presented in
Figs.~\ref{fig:crossover} and ~\ref{fig:single_poly} were measured
in a vibrating sample magnetometer (VSM) at 10\,K, while ramping
the field with 0.6\,T\,min$^{-1}$. All samples were checked by
magnetoscanning \cite{Fug07} to ensure homogeneous properties.

The samples were irradiated in the central irradiation facility of
the TRIGA MARK II reactor in Vienna at a power of 250\,kW
(thermal/fast neutron flux density: $6/7.6\times
10^{16}\,m^{-2}\,s^{-1}$). The samples were sealed into a quartz
tube to avoid contact between the samples and the cooling water in
the reactor. The quartz tubes were put into aluminium containers,
which were filled with water and placed at the correct irradiation
position. The temperature during irradiation did not exceed 60
$^\circ$C. All fluences refer to fast neutrons ($E>0.1$\,MeV), the
corresponding thermal neutron fluences ($E< 0.55$\,eV) were
smaller by a factor of 1.27 \cite{Web86}. Neutron irradiation
creates a wide spectrum of defects in the superconductor by
elastic collisions of the neutrons with the atoms. Depending on
the kinetic energy transferred from the neutron to the primary
recoil atom, point defects or larger defect structures appear. In
an avalanche like process a huge number of atoms can be displaced,
leading to local melting of the crystal lattice. The result is a
spherical region of amorphous material with a mean diameter of
about 3 to 6\,nm \cite{Fri93}. While fast neutrons ($E>0.1$\,MeV)
transfer sufficient energy to the primary recoil atom to produce
collision cascades, thermal neutrons ($E<0.55$\,eV) do not lead to
defects in the material because their energy is below the
displacement threshold. Epithermal neutrons (in the keV range)
lead to the formation of point defects and point defect clusters.
The density of collision cascades after irradiation to
$10^{22}$\,m$^{-2}$ is about\cite{Fri93} $5\times
10^{22}$\,m$^{-3}$, which corresponds to a mean distance between
two cascades of 27\,nm, the lattice parameter of the flux line
lattice at 3.2\,T.

\section{Results and Discussion}
\subsection{Transition Temperature}
The transition temperature of the MODRAB samples was 90.6\,K, of
the PLDYSZ 88.5\,K and of the MOCVDMgO 89.1\,K. Neutron
irradiation decreases the transition temperature, \Tc, which is
ascribed to enhanced impurity scattering caused by the introduced
defects. Non-magnetic impurities do not reduce \Tc\ in isotropic
s-wave superconductors, but in anisotropic and d-wave
superconductors \cite{Mil88,Rad93}. The effect is comparatively
small at the chosen fluences. We find a linear decrease with
neutron fluence by around $2\times10^{-22}$\,Km$^2$ (2 \,K at a
fluence of $10^{22}$\,m$^{-2}$), which is around half that found
in single crystals \cite{Sau91,Wer00}. The transitions normally do
not broaden after irradiation to these neutron fluences.
\subsection{Irreversibility line}
\begin{figure}
\includegraphics[width=0.5\textwidth]{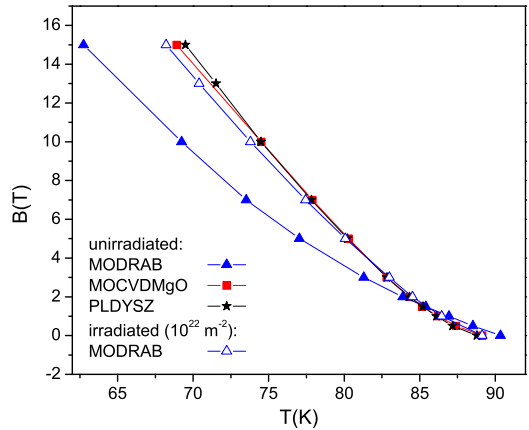}
\caption{\label{fig:IL}
  The irreversibility fields ($H\parallel c$) of the PLDYSZ and MOCVDMgO samples are nearly
  identical and higher than those of the MODRAB sample.
  All irreversibility lines become similar after neutron irradiaton, since only that of the MODRAB sample is changed significantly.}
\end{figure}
The irreversibility lines ($H\parallel c$) of the samples are
compared in Fig.~\ref{fig:IL}. The PLDYSZ and MOCVDMgO samples
behave almost identically, the irreversibility fields of the
MODRAB samples are lower (except near \Tc), but shift close to the
values of the other samples after irradiation to
10$^{22}$\,m$^{-2}$. The latter do not change significantly in the
accessible field range, since the moderate decrease in \Tc, which
somewhat decreases \Birr\ at high temperatures, is counterbalanced
by a slight steepening of the irreversibility line. We note that
neutron irradiation cannot improve the irreversibility line of
state-of-the-art coated conductors and that all samples behave
similarly after irradiation. It seems that the MODRAB samples have
a comparatively low density of pinning efficient defects, which is
increased by the irradiation, thus enhancing \Birr. The
irreversibility line becomes quite insensitive to the defect
concentration at high defect densities, which could indicate an
upper limit for \Birr \cite{Fig06}.

The irradiation has no beneficial effect on the irreversibility
line for the other main field orientation ($H\parallel ab$) and we
observe no significant effect (MODRAB) or a slight degradation
(other samples). The variation of the intrinsic pinning potential
is on the length scale of the c-axis lattice parameter (equivalent
to a very high density of pinning centers) and presumably much
more efficient than the cascades due to its correlated two
dimensional character.
\subsection{Critical Currents}
\begin{figure}%
\includegraphics[width=.5\textwidth]{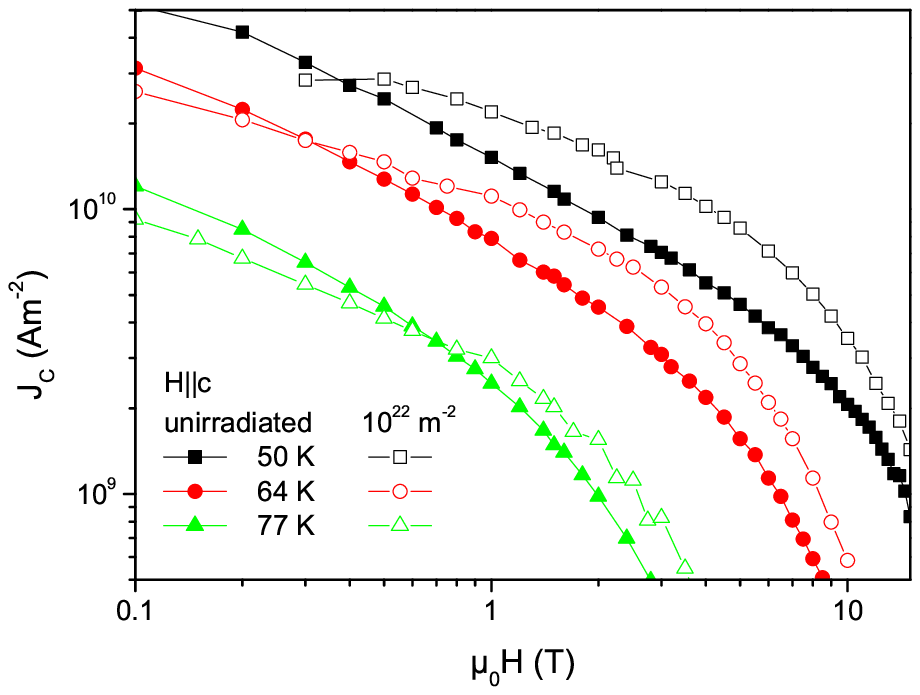}%
\includegraphics[width=.5\textwidth]{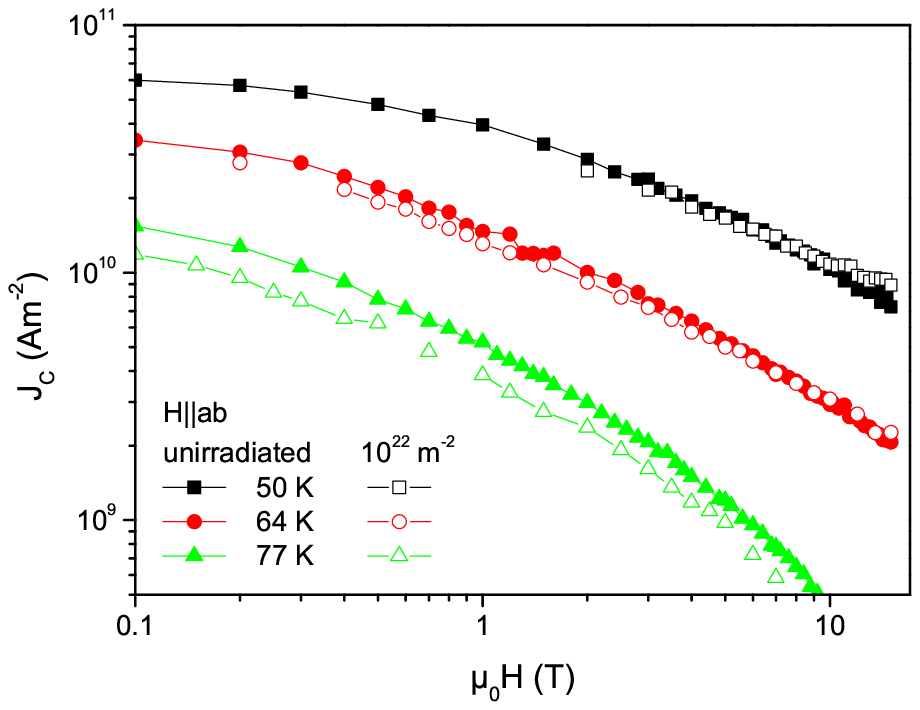}%
\caption{\label{fig:Jc}
  Neutron irradiation enhances the critical current densities in a large field
  and temperature range, but decreases them at low fields or close to \Tc.
  The positive effect is larger if the field is oriented perpendicular to the tape
  (left panel) than in the parallel orientation (right panel).}
\end{figure}
Depending on field, temperature, fluence, and the original defect
structure, the irradiation will enhance or decrease the critical
currents. The strongest increase was observed in the sample with
the weakest as grown pinning structure (MODRAB). The corresponding
data at 77\,K, 64\,K, and 50\,K are presented in Fig.~\ref{fig:Jc}
for both main field orientations. The critical currents decrease
at low magnetic fields after irradiation, which is discussed in
detail in the next section. \Jc$(B)$ becomes flatter after the
irradiation at low and intermediate fields, which leads to a
crossover and an enhancement at intermediate and high fields. The
flattening manifests itself in a smaller exponent $\alpha$ of the
power-law $\Jc\propto B^{-\alpha}$, which is a reasonable
approximation for the field dependence of \Jc\ at intermediate
fields. Note that the crossover field between the enhancement and
degradation of \Jc\ is generally higher for the field parallel to
the ab planes (right panel in Fig.~\ref{fig:Jc}), where an
enhancement is found only at high fields and low temperatures. The
currents degrade in the whole field range at 77\,K. The (relative)
enhancement is  more pronounced at low temperatures for both
orientations, which is in contrast to earlier findings on single
crystals \cite{Sau91,Sau98}. This is a consequence of the
competition between pinning and thermal energy, which reduces the
efficiency of comparatively small defects at high temperatures.
The collision cascades are obviously the largest pinning centers
in single crystals, but this does not seem to be the case in
coated conductors, where a large number and variety of linear and
planar defects exists \cite{Fol07}, which are more efficient than
the cascades at high temperatures. In addition, neutron
irradiation also introduces smaller defects, which are efficient
only at low temperatures.

The other types of samples (PLDYSZ, MOCVDMgO) behave qualitatively
similarly but data are available only for a fluence of $2\times
10^{21}$\,m$^{-2}$ so far. A similar reduction of \Jc\ at low
fields turns to an enhancement at comparable crossover fields,
which hardly depend on the neutron fluence (see
Fig.~\ref{fig:crossover}). The field dependence becomes again
weaker at intermediate field, which reduces $\alpha$. Only at high
fields, the hardly changed irreversibility field causes a
difference to the MODRAB samples. While the relative enhancement
diverges for $H\parallel c$ at the originally lower
irreversibility fields in the MODRAB samples, the \Jc($B$) curves
merge again near the hardly changed irreversibility fields in the
PLDSZ and MOCVDMgO samples. At high temperatures, where the
irreversibility field is slightly reduced due to the decrease of
\Tc, even a second crossover is observed, and \Jc\ is only
enhanced at intermediate fields.
\subsection{Granularity effects}
\begin{figure}
\includegraphics[width=0.5\textwidth]{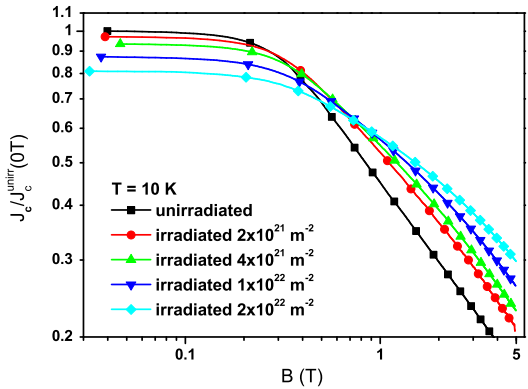}
\caption{\label{fig:crossover}
  Crossover in \JcB{} between the irradiated and the unirradiated state
  after sequential irradiation (sample MODRAB).}
\end{figure}

\begin{figure}%
\includegraphics[width=.5\textwidth]{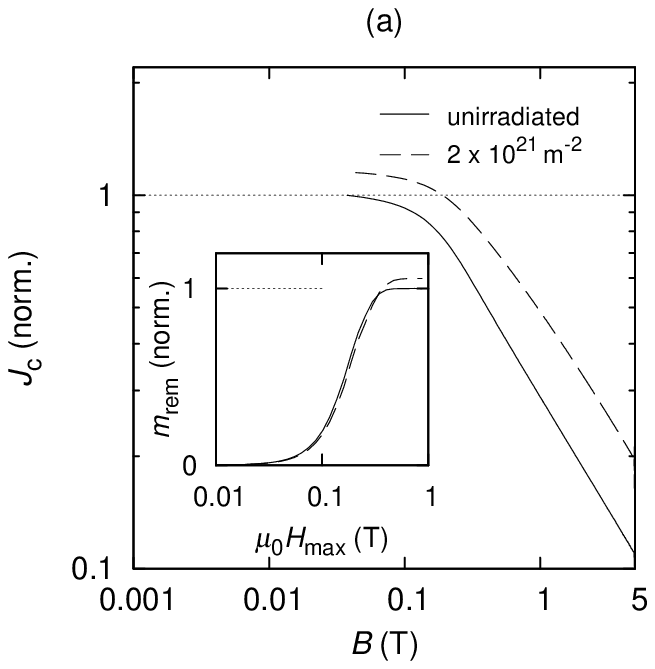}%
\includegraphics[width=.5\textwidth]{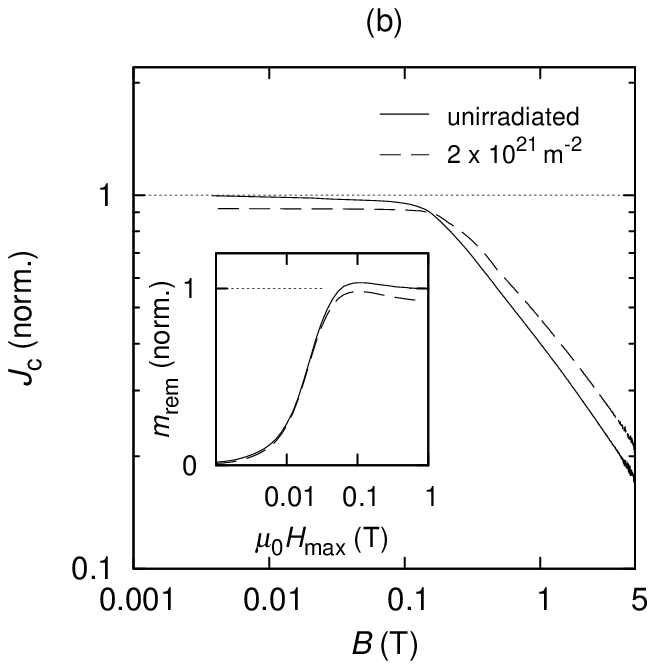}%
\caption{\label{fig:single_poly}
  Comparison of the effect of neutron irradiation
  on samples grown on single-crystalline (a)
  and poly-crystalline (b) substrates.
  The critical current density in the single-crystalline sample
  is enhanced in the entire field range,
  while the granular sample exhibits a crossover
  at about the same field as state-of-the-art coated conductors.
  The inset shows that the remnant moment of the granular sample
  decreases when the grain boundary currents are saturated
  and the flux penetrates the grains.
  (Data are normalised to the unirradiated state.)}
\end{figure}

The crossover in the field dependence \JcB{} between the
unirradiated state and the irradiated state
(Fig.~\ref{fig:crossover}), is a feature that is
generally observed on conductors grown by physical as well as chemical deposition. 
The twofold effect of the irradiation on the current transport
indicates that two regimes must be separated: a high field region
(\unit{> 1}{T}) which benefits from the additional pinning centres
and a low field domain, where a different mechanism causes a
\Jc{}-depression.
This is explained by considering the specific nature of current
transport in a granular medium, i.e., the critical current density
in the grains \JcG{} has to be distinguished from the grain
boundary critical current density \JcGB{} flowing across the
sample. Numerous experiments on bicrystals \cite{Hil02} showed
that, depending on the grain boundary misorientation angle, the
current transport is controlled by \JcGB{} at low fields, whereas
at fields in the Tesla range \JcG{} limits the current transport.
Although it is not fully established that the behaviour of a
single well-defined grain boundary is representative for the
complex grain boundary network of a coated conductor
\cite{Fel08,Dur09}, the current limitation by the grains and an
increase of \JcG{} after irradiation are straightforward
explanations for the observed \Jc{} improvement in high fields.
Further evidence for this scenario is provided by neutron
irradiation of YBCO thin films grown on single-crystalline
substrates, which can be regarded as single macroscopic grains and
thus serve as a model system for the individual microscopic grains
in a coated conductor. These films show, in agreement with the
above assumption of an increase in \JcG{}, enhanced critical
current densities in the entire field range (cf.
Fig.~\ref{fig:single_poly}a and reference~\onlinecite{Hen07}), if
the temperature is not too close to the superconducting transition
temperature , i.e., when the (moderate) reduction of \Tc{} after
irradiation is negligible.
The fact that the \Jc{} decrease occurs at rather low fields,
where \JcGB{} is dominant in the bicrystal experiments, suggests
that the grain boundary network is responsible for this effect.
Two different but not mutually exclusive mechanisms are able to
explain the reduction of \JcGB{}. First, it is important to
notice, that the density of the irradiation defects is too low to
reduce the cross-section for current transport significantly. But
the defects are mobile during the irradiation and the grain
boundaries represent natural barriers for their movement,
which causes the defects to accumulate---%
the grain boundaries deteriorate after irradiation. This effect
has, however, only an impact on the overall \Jc{}, if the current
transport is limited by the grain boundaries. Hence, the crossover
indicates the transition between the low field regime, which is
dominated by the grain boundary network, and the high-field regime
controlled by the properties of the grains.

Another explanation for the \Jc{} reduction at low fields is based
on the field exerted by the intra-granular shielding currents on
the grain boundaries: If \JcG{} is larger than \JcGB{}, additional
flux remains trapped in the grains after a change of the external
applied field and the circulating currents in the grain generate
an additional field at the grain boundaries. Thus, the enhancement
of \JcG{} by the irradiation, which is responsible for the \Jc{}
improvement at high fields, increases the grain boundary field and
thereby reduces the field dependent \JcGB{} at low applied fields.
An experimental test is provided by a method similar to
Ref.~\onlinecite{Mul94}, where the remnant magnetic moment \mrem{}
of the sample is assessed after applying successively larger
fields \Hmax. If \JcGB{} is smaller than \JcG{}, flux will
initially penetrate
along the grain boundaries. 
After the grain boundary network has saturated and additional flux
is trapped only in the grains, $\mrem(\Hmax)$ decreases as the
intra-grain currents increase the field at the grain boundary.
\footnote{Note, 
that the contribution of the currents in the grains to \mrem{} is
not significant, because $m\propto r^3$, where $r$ is the radius,
which is a few millimetres for the sample and more than a hundred
times less for a single grain. Thus, the dominating component of
the total magnetic moment stems from the currents flowing across
the grain boundaries in all samples expect those with high-angle
grain boundaries, which makes a direct assessment of \JcG{} and
\JcGB{} (as for example in Ref.~\onlinecite{Ton02}) impossible.}
The expected behaviour of $\mrem(\Hmax)$ is clearly evident in
samples grown at an early stage of optimisation
(Fig.~\ref{fig:single_poly}b) to the granular substrate and thus
carry a rather low \JcGB{} \cite{Hen07}. Despite the smaller \Jc{}
of this sample, the crossover occurs at fields comparable to
state-of-the-art coated conductors.
Although the two proposed mechanisms are different, they do not
exclude each other, but rely on the same argument: a transition
between grain boundary to grain controlled current transport
leading to a crossover in \JcB{} between the irradiated and the
unirradiated state. It should further be noted that the complexity
of the grain boundary network does not allow to draw clear
distinctions and the crossover indicates only approximately the
transition from one regime to the other.

\subsection{Angular dependence of \Jc}
\begin{figure}
\includegraphics[width=0.5\textwidth]{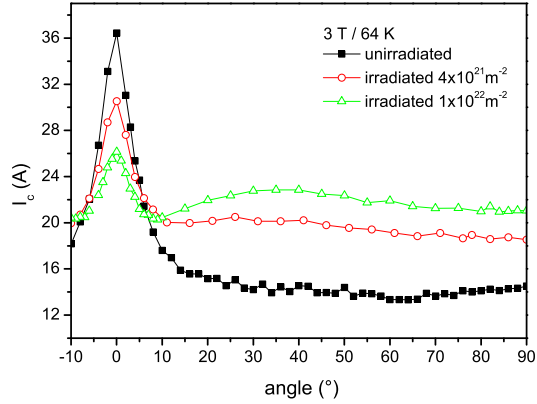}
\caption{\label{fig:angamsc} The irradiation reduces the \Jc-
anisotropy (sample MODRAB). The minimum in \Jc\ appears near
9$^\circ$, where the currents are hardly changed by the introduced
defects.}
\end{figure}
The angular dependence of the critical current in the MODRAB
sample at 64\,K and 3\,T is presented in Fig.~\ref{fig:angamsc}.
The pronounced peak at 0$^\circ$ either comes from intrinsic
pinning by the periodic variation of the order parameter along the
c-axis, or from correlated defects, such as stacking faults. There
is hardly any trace of a peak at 90$^\circ$ in the unirradiated
tape. This indicates that the density of c-axis correlated defects
is small. The random, uncorrelated disorder introduced by the
neutrons reduces the \Ic-anisotropy,
$\gamma_{I_\mathrm{c}}=I_\mathrm{c}^\mathrm{max}/I_\mathrm{c}^\mathrm{min}$,
from 2.7 to 1.3 at a fluence of $10^{22}$\,m$^{-2}$. The latter is
considerably smaller than the \Ic-anisotropy in a thin film with a
claimed (intrinsic or effective) anisotropy of only\cite{Gut07}
1.5, obtained by a scaling analysis \Jc($\theta$) \cite{Gut07} on
the basis of the ansiotropic scaling approach by Blatter et al.
\cite{Bla92} Such a small intrinsic anisotropy can be excluded in
the present sample, since the anisotropy of the irreversibility
fields is much higher and not significantly altered by the
irradiation. The shape of \Jc($\theta$) in the irradiated state is
extremely interesting. The peak at 0$^\circ$ is followed by a
minimum at about 9$^\circ$ and another broad maximum centered at
around $\sim 35^\circ$. Note that this behaviour was induced by
the addition of random, uncorrelated disorder and is hardly
compatible with the anisotropic scaling approach, although this
model assumes such a defect structure, but with weak instead of
strong pinning defects. This is not unexpected, since the scaling
of \Jc\ is based on single-vortex weak-collective pinning
\cite{Bla92} which predicts extremely small currents and is
obviously not applicable to ``high-\Jc'' coated conductors.

\begin{figure}
\includegraphics[width=0.5\textwidth]{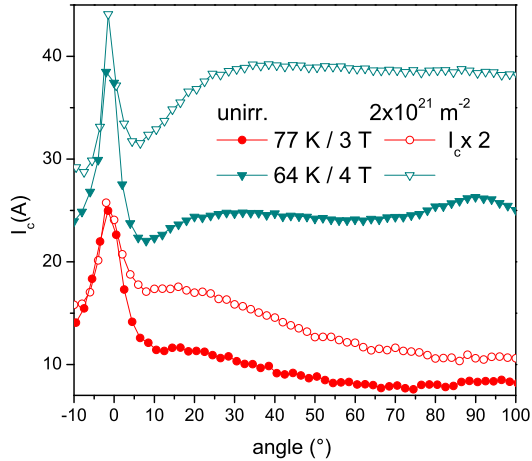}
\caption{\label{fig:angsp} The peak at 90$^\circ$ indicates the
presence of c-axis correlated defects in the MOCVDMgO sample. It
disappears after neutron irradiation, the minimum at $\sim
9^\circ$ is amplified.}
\end{figure}

The MOCVDMgO samples contain c-axis correlated defects, as
indicated by the peak at around 90$^\circ$ shown in
Fig.~\ref{fig:angsp}. This peak disappears after irradiation and
the currents hardly change between 25$^\circ$ and 90$^\circ$ at
64\,K and 4\,T, in contradiction to the anisotropic scaling
approach. The minimum, again at about 9$^\circ$, which was present
in these samples also before irradition, is amplified. Only the
intrinsic pinning remains practically unaffected.

At higher fields (not shown in Fig.~\ref{fig:angsp}), a minimum
occurs at 90$^\circ$, which is an obvious consequence of the small
irreversibility field in this orientation. The ratio of the
irreversiblity fields in the main field orientations is consistent
with anisotropic scaling and with expectations for the intrinsic
anisotropy ($B_\mathrm{irr}^{ab}/B_\mathrm{irr}^{c}\sim 4$) and
seems to represent a much more reliable estimate for the intrinsic
anisotropy (the scaling of fields is independent of pinning
models) than the angular dependence of \Jc, since this approach
does not work for the relevant pinning centers in coated
conductors.

\subsection{Coated conductors for fusion magnets}
\begin{figure}
\includegraphics[width=0.5\textwidth]{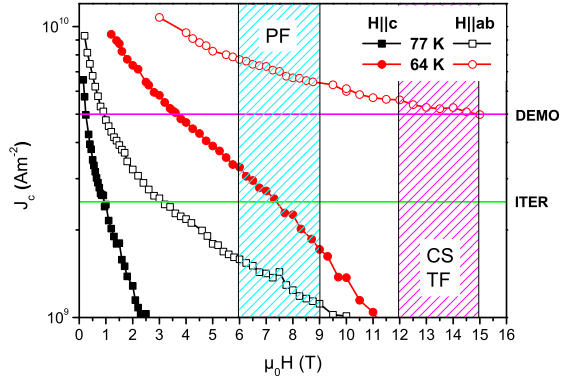}
\caption{\label{fig:fusion}Commercial state-of-the-art conductors
fulfill the requirements of fusion magnets on \Jc\ at 64\,K, a
temperature, where liquid nitrogen (at low pressure) can be used
as coolant, if the magnetic field is parallel to the ab planes.
For the other main field orientation as well as at 77\,K the
performance is still too low.}
\end{figure}
Powerful magnets will be needed to confine the plasma of future
fusion power plants. The use of high temperature superconductors
instead of Nb$_3$Sn would allow operation at higher temperatures
and/or fields, which would reduce the cooling requirements and
could also result in a more compact and cheaper design. However,
only first concepts for fusion magnets made of coated conductors
are being developed at present. Cables with a current capability
in the ten kA range are needed and the strands have to be somehow
twisted \cite{Gol09} in order to reduce ac losses. Electrical,
mechanical, and thermal stabilization of such wires and quench
detection/protection will be demanding issues. Apart from these
technological issues, the superconducting properties have to
fulfill two basic requirements. The critical current densities
must be comparatively high, since the superconducting volume
fraction of a coated conductor is low (typically 1-2\%) and they
must not degrade during operation in a hard radiation environment
(fast neutrons). A rough estimate of the required intrinsic
critical current density leads to about $2.5\times
10^9$\,Am$^{-2}$ ($5\times 10^9$\,Am$^{-2}$) at about 12\,T
(14\,T) \cite{Fug09}, assuming the overall current density of the
actual ITER (DEMO) cable design for the toroidal field (TF) coils
or the central solenoid (CS). These values are certainly
achievable with present conductors, but define the operation
temperature. Cooling with liquid nitrogen would be favorable, but
is possible only above 64\,K (at low pressure).
Figure~\ref{fig:fusion} demonstrates that this seems feasible
(although not at 77\,K) with present conductors (MOCVDMgO) if the
field is aligned with the ab planes. The magnetic field in the
fusion coils will be close to this orientation, but, depending on
the cable and coil design, orthogonal components will occur. In
the other limiting case ($H\parallel c$), the performance
currently fails significantly. Since the currents drop quickly at
small misalignment angles (Figs.~\ref{fig:angamsc} and
\ref{fig:angsp}), further conductor development seems mandatory in
order to reach the required current densities at these high
temperatures. The field in the poloidal field (PF) coils is much
smaller (up to about 6\,T in ITER and 9\,T in DEMO), which is also
highlighted in Fig.~\ref{fig:fusion}. Even these requirements seem
to be demanding at 77\,K or for $H\parallel c$. The expected life
time fluence (fast neutrons) at the main fusion magnets (PF, CS)
is around $2\times 10^{22}$\,m$^{-2}$ and might be even higher in
future fusion power plants. Since the currents at 77\,K already
degrade at a fluence of $10^{22}$\,m$^{-2}$, radiation resistance
is an important issue for the application of coated conductors in
fusion magnets, at least at elevated temperatures.

\section{Conclusions}
The influence of fast neutron irradiation on differently
fabricated coated conductors was found to be similar. The critical
currents decrease at low magnetic fields, where inter-granular
currents set the upper limit for loss free current flow. This
decrease can be understood either by a slight degradation of the
grain boundaries, or by the increasing shielding currents within
the grains, which generate an additional field at the grain
boundaries.

The irradiation enhances \Jc\ at intermediate and high magnetic
fields, where the macroscopic current is limited by pinning, which
is improved by the additional pinning centers. Although these
pinning centers are uncorrelated, the angular dependence of \Jc\
does not obey anisotropic scaling. This is a direct consequence of
the pinning strength of these defects, which cannot be considered
as weak, as assumed in the scaling approach. Since this should be
true for any relevant pinning centers in these ``high-\Jc''
conductors, anisotropic scaling based on weak collective single
vortex pinning is not applicable.

The irreversibility line in the best available YBCO coated
conductors seems to be close to optimum, because a higher
concentration of strong defects does not lead to further
improvements.

\begin{acknowledgments}

This work, supported by the European Communities under the
contract of Association between EURATOM/OEAW was partly carried
out within the framework of the European Fusion Development
Agreement. The views and opinions expressed herein do not
necessarily reflect those of the European Commission. One of us
(M.C.) wishes to acknowledge financial support through the NESPA
project.

\end{acknowledgments}

\end{document}